\documentclass{article}
\usepackage{amssymb}
\begin{document}
\title{Solution of the Schr\"odinger equation making use of time-dependent constants of motion}

\author{G.F.\ Torres del Castillo \\ Departamento de F\'isica Matem\'atica, Instituto de Ciencias \\
Universidad Aut\'onoma de Puebla, 72570 Puebla, Pue., M\'exico}

\maketitle

\begin{abstract}
It is shown that if a complete set of mutually commuting operators is formed by constants of motion, then, up to a factor that only depends on the time, each common eigenfunction of such operators is a solution of the Schr\"odinger equation. In particular, the operators representing the initial values of the Cartesian coordinates of a particle are constants of motion that commute with each other and from their common eigenfunction one readily obtains the Green function.
\end{abstract}

\noindent PACS numbers: 03.65.-w

\section{Introduction}
In the standard procedure to solve the Schr\"odinger equation for a time-independent Hamiltonian one makes use of the method of separation of variables, or one starts looking for a set of mutually commuting operators that also commute with the Hamiltonian. In fact, the separable solutions of the Schr\"odinger equation are common eigenfunctions of a set of mutually commuting operators, with the separation constants being the eigenvalues of such operators. Usually, these operators do not depend explicitly on the time and, since they commute with the Hamiltonian, are constants of motion.

As we shall show below, there is no reason to restrict ourselves to time-independent operators; we can find solutions of the Schr\"odinger equation that are common eigenfunctions of a set of, possibly time-dependent, mutually commuting operators that are constants of motion (and, therefore, may not commute with the Hamiltonian). This method is analogous to that given by the Liouville theorem on the solutions of the Hamilton--Jacobi equation (see, e.g., Ref.\ \cite{Ge}).

In Section 2 we establish the basic results of this paper and in Section 3 we present several examples, exposing the advantages of the method. We show that making use of the operators that represent the initial position of a particle one readily obtains the Green function of the corresponding Hamiltonian.

\section{Basic results}
Let $A$ be a, possibly time-dependent, Hermitean operator that is a constant of motion, i.e.,
\begin{equation}
{\rm i} \hbar \frac{\partial A}{\partial t} + [A, H] = 0, \label{cons}
\end{equation}
where $H$ is the Hamiltonian of the system under consideration. If $\psi$ is an eigenfunction of $A$ with eigenvalue $\lambda$,
\begin{equation}
A \psi = \lambda \psi \label{eig}
\end{equation}
and $\psi$ satisfies the Schr\"odinger equation
\begin{equation}
{\rm i} \hbar \frac{\partial \psi}{\partial t} = H \psi, \label{sch}
\end{equation}
then $\lambda$ is a constant. Indeed, taking the derivative with respect to the time of both sides of (\ref{eig}), we obtain
\[
\left( {\rm i} \hbar \frac{\partial A}{\partial t} \right) \psi + A \left( {\rm i} \hbar \frac{\partial\psi}{\partial t} \right) = {\rm i} \hbar \frac{{\rm d} \lambda}{{\rm d} t} \psi + \lambda {\rm i} \hbar \frac{\partial \psi}{\partial t}
\]
and, making use of Eqs.\ (\ref{cons}) and (\ref{sch}),
\[
- [A, H] \psi + AH \psi = {\rm i} \hbar \frac{{\rm d} \lambda}{{\rm d} t} \psi + \lambda H \psi,
\]
that is,
\[
HA \psi = {\rm i} \hbar \frac{{\rm d} \lambda}{{\rm d} t} \psi + \lambda H \psi
\]
and using Eq.\ (\ref{eig}) again, it follows that ${\rm d} \lambda/{\rm d} t = 0$.

Conversely, assuming that $A$ is a constant of motion and that its eigenvalues are constant, from (\ref{eig}) we have
\[
\left( {\rm i} \hbar \frac{\partial A}{\partial t} \right) \psi + A \left( {\rm i} \hbar \frac{\partial\psi}{\partial t} \right) = \lambda {\rm i} \hbar \frac{\partial \psi}{\partial t}
\]
or, by virtue of (\ref{cons}),
\[
- [A, H] \psi + A \left( {\rm i} \hbar \frac{\partial\psi}{\partial t} \right) = \lambda {\rm i} \hbar \frac{\partial \psi}{\partial t},
\]
which amounts to
\[
- AH \psi + H \lambda \psi +  A \left( {\rm i} \hbar \frac{\partial\psi}{\partial t} \right) = \lambda {\rm i} \hbar \frac{\partial \psi}{\partial t},
\]
or
\[
A \left( {\rm i} \hbar \frac{\partial\psi}{\partial t} - H \psi \right) = \lambda \left( {\rm i} \hbar \frac{\partial\psi}{\partial t} - H \psi \right).
\]
If the spectrum of $A$ is non-degenerate, it follows that
\[
{\rm i} \hbar \frac{\partial\psi}{\partial t} - H \psi = \mu \psi,
\]
where $\mu$ is some complex-valued function of $t$ only. Then, if $F(t)$ is a solution of
\begin{equation}
\mu F + {\rm i} \hbar \frac{{\rm d} F}{{\rm d} t} = 0, \label{ode}
\end{equation}
we find that $F \psi$ satisfies the Schr\"odinger equation. In fact,
\begin{eqnarray*}
{\rm i} \hbar \frac{\partial (F \psi)}{\partial t} & = & F (H \psi + \mu \psi) + \left( {\rm i} \hbar \frac{{\rm d} F}{{\rm d} t} \right) \psi \\
& = & H(F \psi) + \left( \mu F + {\rm i} \hbar \frac{{\rm d} F}{{\rm d} t} \right) \psi.
\end{eqnarray*}
(Note that, if $\psi$ is an eigenfunction of $A$ with eigenvalue $\lambda$, then, if $F(t)$ is a function of $t$ only, $F(t) \psi$ is also an eigenfunction of $A$ with eigenvalue $\lambda$.) When the spectrum is degenerate, the proof is similar, considering linear combinations of the eigenfunctions of $A$ with the same eigenvalue $\lambda$ and, in place of (\ref{ode}), one would have to solve a linear system of first-order differential equations, which may be as complicated as the original Schr\"odinger equation. However, if we find a complete set of mutually commuting operators that are constants of motion (complete in the sense that, up to a factor, there is only one common eigenvector of these operators for a given set of the eigenvalues), the common eigenfunctions of such a set of operators can be chosen in such a way that they are solutions of the Schr\"odinger equation (assuming that the eigenvalues are constant).

This result is analogous to the Liouville theorem of the Hamiltonian mechanics, according to which if we have a set of constants of motion such that their Poisson brackets are equal to zero, they can be used to find complete solutions of the Hamilton--Jacobi equation.

\section{Examples}
In this section we give several examples, finding solutions of the Schr\"odinger equation starting from the eigenfunctions of constants of motion that depend explicitly on the time.

\subsection{The one-dimensional harmonic oscillator}
One can readily verify that the operator
\begin{equation}
X_{0} \equiv x \cos \omega t - \frac{p}{m \omega} \sin \omega t \label{xzero}
\end{equation}
is a constant of motion if the Hamiltonian is given by
\begin{equation}
H = \frac{p^{2}}{2m} + \frac{m \omega^{2}}{2} x^{2}. \label{hoh}
\end{equation}
The eigenvalue equation $X_{0} \psi = x_{0} \psi$ amounts to the {\em first-order}\/ differential equation
\[
\left( x \cos \omega t - \frac{\hbar \sin \omega t}{{\rm i} m \omega} \frac{\partial}{\partial x} \right) \psi = x_{0} \psi
\]
whose solution is readily found to be
\begin{equation}
\psi = F(t) \exp \left[ \frac{{\rm i} m \omega}{2 \hbar \sin \omega t} \big( x^{2} \cos \omega t - 2 x_{0} x \big) \right], \label{wf1}
\end{equation}
where $F(t)$ is a function of $t$ only.

Assuming that $x_{0}$ is a constant, one finds that the wavefunction (\ref{wf1}) satisfies the Schr\"odinger equation, ${\rm i} \hbar \partial \psi/\partial t = H \psi$, if and only if [cf.\ Eq.\ (\ref{ode})]
\[
\frac{{\rm d} \ln F}{{\rm d} t} = - \frac{\omega \cos \omega t}{2 \sin \omega t} + \frac{m \omega^{2} x_{0}{}^{2}}{2 {\rm i} \hbar \sin^{2} \omega t},
\]
hence
\[
F = \frac{C}{\sqrt{\sin \omega t}} \exp \left( \frac{{\rm i} m \omega x_{0}{}^{2} \cos \omega t}{2 \hbar \sin \omega t} \right),
\]
where $C$ is a constant, and substituting this expression into (\ref{wf1}) we obtain
\begin{equation}
\psi_{x_{0}}(x,t) = \frac{C}{\sqrt{\sin \omega t}} \exp \left\{ \frac{{\rm i} m \omega}{2 \hbar \sin \omega t} \big[ (x^{2} + x_{0}{}^{2}) \cos \omega t - 2 x_{0} x \big] \right\}. \label{wf2}
\end{equation}
As usual, the wavefunctions $\psi_{x_{0}}$ are orthogonal for different values of the eigenvalue $x_{0}$. Note also that the functions (\ref{wf2}) are not separable.

In the present example, $H$ is time-independent and, therefore, there is a set of stationary states, $\phi_{n}(x) \exp (- {\rm i} E_{n} t/\hbar)$, where the $E_{n}$ are the eigenvalues of $H$. Hence, the wavefunction (\ref{wf2}) must be expressible in the form
\[
\psi_{x_{0}}(x,t) = \sum_{n} c_{n}(x_{0}) \phi_{n}(x) \exp \left( - \frac{{\rm i} E_{n} t}{\hbar} \right)
\]
and the symmetry of the right-hand side of (\ref{wf2}) under the interchange of $x$ and $x_{0}$ implies that
\[
\psi_{x_{0}}(x,t) = \sum_{n} \phi_{n}(x_{0}) \phi_{n}(x) \exp \left( - \frac{{\rm i} E_{n} t}{\hbar} \right)
\]
i.e., with the appropriate value of the constant $C$, $\psi_{x_{0}}(x,t)$ is the time-dependent {\em Green function}\/ of the one-dimensional harmonic oscillator (cf., for instance, Refs.\ \cite{MK,FH,S}). This was to be expected since the operator $X_{0}$ is, in the Heisenberg picture, the position operator at $t = 0$ and, therefore, $\psi_{x_{0}}$ corresponds to the state of the particle with a well defined value (equal to $x_{0}$) of the position at $t = 0$. The value of the normalization constant $C$ can be determined by the condition
\begin{equation}
\lim_{t \rightarrow 0} \int_{- \infty}^{\infty} \psi_{0}(x,t) \, {\rm d} x = 1, \label{nor}
\end{equation}
which gives
\[
C = \sqrt{\frac{m \omega}{2 \pi {\rm i} \hbar}}.
\]

As is well known \cite{FH}, from the Green function (\ref{wf2}) one can obtain the expression of the stationary states $\phi_{n}(x) \exp (- {\rm i} E_{n} t/\hbar)$ in terms of the Hermite polynomials.

\subsection{A time-dependent Hamiltonian}
Now we shall consider the time-dependent Hamiltonian
\begin{equation}
H = \frac{p^{2}}{2m} - ktx, \label{tdh}
\end{equation}
where $k$ is a constant. The operator
\begin{equation}
X_{0} \equiv x - \frac{tp}{m} + \frac{kt^{3}}{3m} \label{x0}
\end{equation}
is a constant of motion and, in the Heisenberg picture, corresponds to the position of the particle at $t = 0$. The eigenfunctions of this operator are determined by the first-order differential equation
\[
\left( x - \frac{\hbar t}{{\rm i} m} \frac{\partial}{\partial x} + \frac{kt^{3}}{3m} \right) \psi = x_{0} \psi,
\]
where $x_{0}$ is the eigenvalue of $X_{0}$. The solutions of this equation have the form
\[
\psi = F(t) \exp \left[ \frac{{\rm i} m}{\hbar t} \left( \frac{x^{2}}{2} - x_{0} x + \frac{kt^{3}x}{3m} \right) \right],
\]
where $F(t)$ is a function of $t$ only. Substituting this expression into the Schr\"odinger equation, assuming that $x_{0}$ is constant, we obtain the equation
\[
\frac{{\rm d} \ln F}{{\rm d} t} = - \frac{1}{2t} + \frac{m}{2 {\rm i} \hbar t^{2}} \left( x_{0}{}^{2} + \frac{4kt^{3}x_{0}}{3m} + \frac{4k^{2}t^{6}}{9m^{2}} \right),
\]
thus,
\begin{equation}
\psi_{x_{0}}(x,t) = \frac{C}{\sqrt{t}} \exp \left\{ \frac{{\rm i} m}{2 \hbar t} \left[ (x - x_{0})^{2} + \frac{2kt^{3} (x - x_{0})}{3m} - \frac{4 k^{2} t^{6}}{45 m^{2}} \right] \right\}, \label{gf2}
\end{equation}
where $C$ is a constant. With the appropriate value of $C$ ($C = \sqrt{m/2 \pi {\rm i} \hbar}$), (\ref{gf2}) is the Green function for the Hamiltonian (\ref{tdh}).

It may be remarked that, since the Hamiltonian (\ref{tdh}) depends explicitly on the time, the corresponding Schr\"odinger equation cannot be solved by separation of variables.

\subsection{Another standard example}
Another example usually considered in connection with the Green functions is that of a particle in a uniform field. If
\begin{equation}
H = \frac{p^{2}}{2m} - eE x, \label{ufh}
\end{equation}
where $e$ and $E$ are constants, one readily finds that
\begin{equation}
X_{0} \equiv x - \frac{tp}{m} + \frac{eEt^{2}}{2m} \label{xc}
\end{equation}
is a constant of motion that corresponds to the position of the particle at $t = 0$. Thus, the eigenvalue equation $X_{0} \psi = x_{0} \psi$ is equivalent to
\[
\left( x - \frac{\hbar t}{{\rm i} m} \frac{\partial}{\partial x} + \frac{eEt^{2}}{2m} \right) \psi = x_{0} \psi,
\]
with the solution
\[
\psi = F(t) \exp \left[ \frac{{\rm i} m}{2 \hbar t} \left( x^{2} - 2x_{0}x + \frac{eEt^{2}x}{m} \right) \right],
\]
where $F(t)$ is a function of $t$ only. Substituting this last expression into the Schr\"odinger equation one obtains the condition
\[
\frac{{\rm d} \ln F}{{\rm d} t} = - \frac{1}{2t} + \frac{m}{2 {\rm i} \hbar t^{2}} \left( x_{0}{}^{2} - \frac{eEt^{2}x_{0}}{m} + \frac{e^{2}E^{2}t^{4}}{4m^{2}} \right)
\]
and, therefore,
\[
\psi_{x_{0}}(x,t) = \frac{C}{\sqrt{t}} \exp \left\{ \frac{{\rm i} m}{2 \hbar t} \left[ (x - x_{0})^{2} + \frac{eEt^{2} (x + x_{0})}{m} - \frac{e^{2}E^{2} t^{4}}{12 m^{2}} \right] \right\},
\]
where $C$ is a constant.

\subsection{A two-dimensional system}
The Hamiltonian
\begin{equation}
H = \frac{1}{2m} \left[ \left( p_{x} + \frac{eB}{2c} y \right)^{2} + \left( p_{y} - \frac{eB}{2c} x \right)^{2} \right] \label{magn}
\end{equation}
corresponds to a charged particle of mass $m$ and electric charge $e$ in a uniform magnetic field $B$. The operators
\begin{eqnarray}
X_{0} & \equiv & \frac{1}{2} (1 + \cos \omega t) x - \frac{1}{2} y \sin \omega t - \frac{1}{m \omega} p_{x} \sin \omega t + \frac{1}{m \omega} (1 - \cos \omega t) p_{y}, \label{2x} \\
Y_{0} & \equiv & \frac{1}{2} (1 + \cos \omega t) y + \frac{1}{2} x \sin \omega t - \frac{1}{m \omega} p_{y} \sin \omega t - \frac{1}{m \omega} (1 - \cos \omega t) p_{x}, \label{2y}
\end{eqnarray}
where $\omega \equiv eB/m c$, are two constants of motion that commute with each other. The operators (\ref{2x}) and (\ref{2y}) correspond to the Cartesian coordinates of the particle at $t = 0$.  (It may be remarked that, for any system, the operators that represent the initial values of the Cartesian coordinates, or of the Cartesian components of the momentum, are constants of motion that commute with each other.)

Since the operators (\ref{2x}) and (\ref{2y}) are linear in the momenta, one readily finds their eigenfunctions. In fact, the eigenfunctions of $X_{0}$ and $Y_{0}$ with eigenvalues $x_{0}$ and $y_{0}$, respectively, are
\begin{eqnarray}
\psi & = & F(t) \exp \left\{ \frac{{\rm i} m \omega}{4 \hbar (1 - \cos \omega t)} \left[ (x^{2} + y^{2} - 2x_{0} x - 2 y_{0} y) \sin \omega t + 2(x_{0} y - y_{0} x) (1 - \cos \omega t) \right] \right\} \nonumber \\
& = & F(t) \exp \left\{ \frac{{\rm i} m \omega}{4 \hbar} \left[ (x^{2} + y^{2} - 2x_{0} x - 2 y_{0} y) \cot {\textstyle \frac{1}{2}} \omega t + 2(x_{0} y - y_{0} x) \right] \right\}, \label{wf2d}
\end{eqnarray}
where $F(t)$ is a function of $t$ only. Substituting (\ref{wf2d}) into the Schr\"odinger equation we obtain the single condition
\[
\frac{{\rm d} \ln F}{{\rm d} t} = - \frac{\omega}{2} \cot {\textstyle \frac{1}{2}} \omega t- \frac{{\rm i} m \omega^{2}}{8 \hbar} (x_{0}{}^{2} + y_{0}{}^{2}) \csc^{2} {\textstyle \frac{1}{2}} \omega t,
\]
therefore,
\begin{equation}
\psi = C \csc {\textstyle \frac{1}{2}} \omega t \exp \left[ \frac{{\rm i} m \omega}{4 \hbar} \left\{ [(x - x_{0})^{2} + (y - y_{0})^{2}] \cot {\textstyle \frac{1}{2}} \omega t + 2(x_{0} y - y_{0} x) \right\} \right], \label{gf2d}
\end{equation}
where $C$ is a constant.

\section{Concluding remarks}
The procedure followed here to find the Green functions highly contrasts with the methods usually employed in the literature (see, e.g., Refs.\ \cite{MK,FH,S}). For instance, the fact that the operators $X_{0}$, defined by Eqs.\ (\ref{xzero}), (\ref{x0}), and (\ref{xc}), are linear in $p$ implies that, in order to find the Green functions, one only has to solve two first-order ordinary differential equations. Even though the results presented in Section 2 also apply in the usual case of time-independent operators that commute with the Hamiltonian, the difference is that there are not many useful time-independent constants of motion linear in the momentum.

\end{document}